\documentstyle[bibnorm]{lamuphys}
\input{psfig}
\begin{document}
\title{Diffractive Physics in the Near Future}

\author{Gilvan A. Alves\inst{1}}

\institute{Centro Brasileiro de Pesquisas F\'{\i}sicas, Rua Xavier Sigaud 150,
22290-180 Rio de Janeiro - RJ, Brazil}

\maketitle

\begin{abstract}
This paper summarize the present status and near term perspectives for 
Diffractive Physics at Tevatron. We describe the 
new detectors that are being installed around the CDF and D\O\ interaction
regions, and discuss the physics topics accessible in view of these
upgrades.

\end{abstract}
\section{Introduction}

Diffraction is certainly one of the oldest subject in physics. In particle 
physics we have seen a big development in the second half of the last century 
\cite{Predazzi}. More recently, with the discovery of Hard Diffraction by the 
UA8-Collaboration \cite{UA8}, we have had a significant development of  
Diffractive Physics both in theoretical and experimental points of view.

On the experimental side we have now many results coming from HERA (H1 and 
ZEUS)\cite{hera} and the Tevatron (CDF and D\O)\cite{alves1} in a large 
number of topics 
including diffractive structure functions, diffractively produced 
jets, diffractive production of Heavy Flavors, etc.

On the theoretical side\cite{lp01}, a great deal of progress have been made in several 
issues, like factorization, diffractive structure functions, Large Rapidity Gap
Survival Probability, etc.
\par
	Although the experimental results from HERA are showing a consistent 
improvement in the quality of data, the Tevatron results are still plagued by 
poor statistics. Not to mention that some processes, like Double Pomeron
Exchange, have not been directly observed, and for most of the data sample, the
kinematical information (t and $\xi$) is still missing.  

Despite of the theoretical progress for the
unification of the soft and hard aspects of the strong interaction, we still need 
more experimental data to guide some of the theoretical development. 
It is important to stress that currently about 40\% of the $p \bar p$ total cross section 
is due to the Pomeron exchange, and at higher energies this value can be even
larger. In this sense, it is extremely important to have precise data
from the physical region of the proton anti-proton interaction at the Tevatron, 
to answer questions like: 
Is the Pomeron picture universal?, what are its hadronic characteristics?, 
Is it  a glueball?, a dual object?, what is the diffractive contribution for 
the heavy flavor cross sections?, can the Higgs and Centauros be produced
diffractively at Tevatron energies? and so on.
\par
	All of these questions lead to the proposal of new devices  
inserted at Tevatron. CDF reinstalled and improved its  Roman Pots 
Spectrometer, and added a new Miniplug Calorimeter plus Beam Shower 
Counters. D\O\ added a new Forward Proton Detector (FPD).

\par
\begin{figure}[htb]
\centerline{\psfig{figure=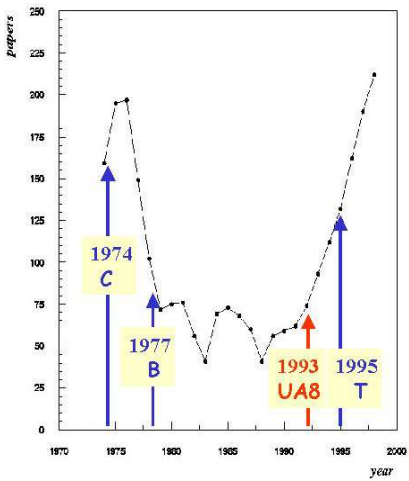,height=2.5in}}
\caption{Number of paper per year in diffractive physics. We 
show four important dates which have some correlation with diffractive physics 
papers. 1974 was the year of charm discovery, three years later, in 1977 the 
bottom, 1993 was the year of the hard diffraction discovery and 1995 the 
top discovery.}
\label{fig:papers}
\end{figure}

	Figure \ref{fig:papers} shows the growing interest in diffractive 
physics. There is a clear growth in the number of papers on this subject in 
particle physics\cite{publicacoes}.
\par	
	
	We are expecting a new period of diffractive results from 
several experiments in the near future. From the Tevatron, the upcoming Run II 
with CDF and D\O; 
From {\it Brookhaven National Laboratory (BNL)}/{\it Relativistic Heavy Ion 
Collider (RHIC)} with the pp2pp experiment, polarized and unpolarized proton - 
proton interactions at energies of 
${\sqrt{s}}$ = 60  -  500 \ GeV; From the {\it Deutsches Elecktronen Synchroton 
(DESY)}, with H1 and ZEUS, the two detectors at HERA, in electron (positron) (27.5 GeV) 
proton (820-920 GeV) interactions. For the far future we will have a new era of 
experiments, with TOTEM \cite{giorgio} being integrated into  {\it Compact 
Muon Solenoid (CMS)} as one of the  detectors of the {\it Large Hadron Collider 
(LHC)}/{\it CERN}, for proton - proton 
at 14 TeV center of mass energy.

\par 

\section{The Tevatron Upgrades for Diffractive Physics CDF and D\O}
	After the end of Run I, the Tevatron accelerator started its upgrade in order 
to achieve higher energy and luminosity. The main goals were to achieve energies up 
to ${\sqrt{s}}$ = 2.0 \ TeV, and the luminosities of $\approx 3\times
10^{32}cm^{-2}s^{-1}$. 
Besides, in order to implement the diffractive program at D\O, some modifications had
to be done to the accelerator itself. 
The two collider detectors CDF and D\O\ were also submitted to several upgrades, but we will 
restrict our discussion here only to those that have direct relation with diffractive physics.

\par
	The Tevatron beam line had to be modified to accommodate the new D\O\ leading
proton spectrometers. This included modifying the electrostatic separator girder, 
extending the cryogenic bypass, removal of a quadrupole magnet ($Q_1$), and 
other small modifications, like drilling a hole on the floor to 
allow the insertion and removal of detectors. 
	
\par
	\subsection{CDF}
	We will summarize the diffractive physics upgrades for CDF. For more 
details, please refer to the paper of Goulianos\cite{goulianos}.
	The CDF upgrades for diffractive physics can be divided in 3 parts:
\begin{enumerate}
	\item{Improve dipole spectrometer (Roman Pots) as is shown in figure
	      \ref{fig:cdfpots}}	

\begin{figure}[htb]
\centerline{\psfig{figure=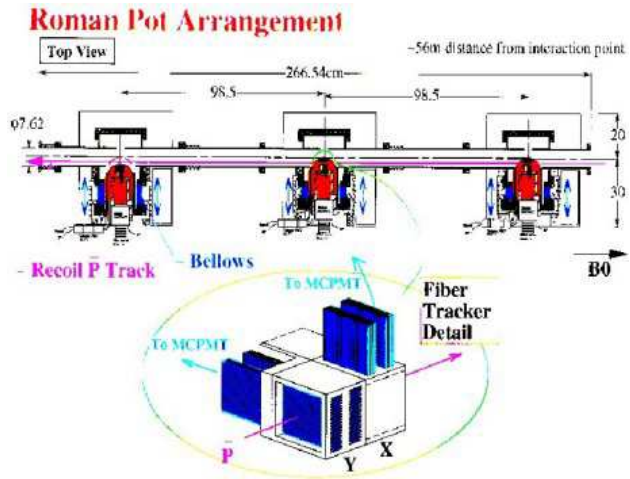,height=2.5in}}
\caption{Roman Pots arrangement of CDF.}
\label{fig:cdfpots}
\end{figure}

	\item{Insertion of a Miniplug Calorimeter for triggering on diffractive
              events. This device, shown in figure \ref{fig:cdfminiplug}, 
              is composed by 50 1/4" thick lead plates 
              corresponding to a total of 2 interaction lengths and $\approx$ 60 
              radiation lengths. 288 signal towers are viewed by 18 Multi Channel 
              PMTs (16 channels each), covering a pseudorapidity region between 
              3.5 and 5.5.}

\begin{figure}[htb]
\centerline{\psfig{figure=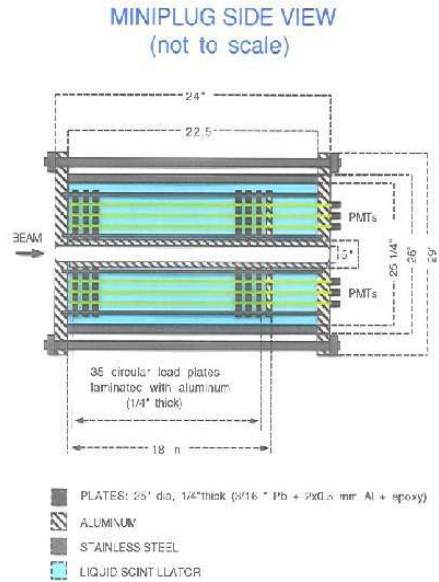,height=3.5in}}
\caption{Side view of the CDF Miniplug Calorimeter.}
\label{fig:cdfminiplug}
\end{figure}
             
	\item{Install new Beam Shower Counters at large rapidity region (5.5 - 7.5). 
            These counters consist of regular plastic scintillators, read out by
            PMTs appropriate for high magnetic fields.}
	    
\end{enumerate}
\subsection{D\O}
\begin{figure}[htb]
\centerline{\psfig{figure=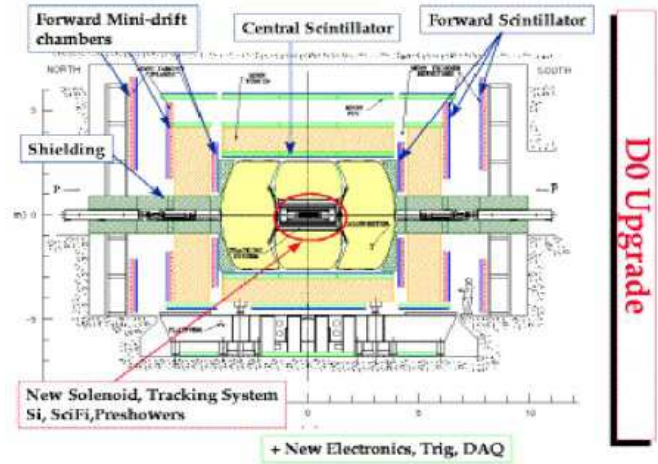,height=2.5in}}
\caption{The main components of the upgraded D\O\ detector for RunII.}
\label{fig:upgrade}
\end{figure}

	The whole D\O\ detector has been submitted to many modifications for
the present run (RunII). Figure \ref{fig:upgrade} summarize the main features.
\par	
\par
	Now we will describe the Forward Proton Detector, the main component of
the D\O\ Diffractive Upgrade.
	\section{Forward Proton Detector}
	The idea of the Forward Proton Detector (FPD) is to cover experimentally a 
large number of topics which will be very important for the progress of  
diffractive physics. More details on the FPD can be read in the reference 
\cite{proposal}.
\par
\begin{figure*}[htb]
\centerline{\psfig{figure=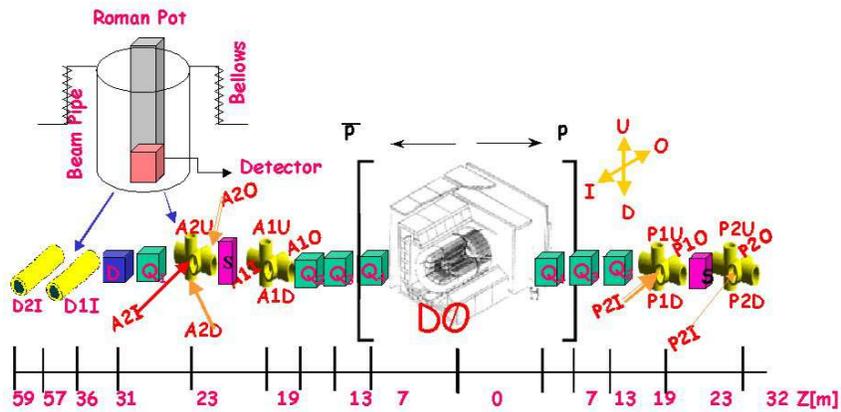,width=4.5in}}
\caption{This figure shows the FPD in the beam line of the Tevatron, in both
sides of the  D\O\ \ detector.}
\label{fig:capa}
\end{figure*}

	The Forward Proton Detector consists of 18 Roman Pots arranged on both 
sides of the D\O\ detector as shown in figure \ref{fig:capa}, which shows the Roman 
Pot locations in the Tevatron beam line. There are two castles\footnote{Structure that
holds the Roman Pots, motors and vacuum equipment.} on the proton 
side indicated by $P1$ \ and  \ $P2$ as shown in figure \ref{fig:capa}. The 
orientation is indicated by the additional letter U for up position, D for down 
position, I for inside position and O for outside position of the pots ($P1U, \ 
P1D, \ P1I, \ P1O$, same notation for $P2$). On the antiproton side we have 
two similar castles, labeled $A1$ \ and \ $A2$ followed by the indication 
of the orientation similar for the proton side. In addition, two others half castles 
on the antiproton side hold the dipole spectrometer\footnote{Named in this way 
for their position after the dipole magnets.} labeled $D1$ \ and \ $D2$. The 
approximated distances of the pots with respect to the interaction point 
(indicated by 0 on the scale) are also shown.
\par	

\par	
\subsection{Roman Pots}
\par
\begin{figure}[htb]
\centerline{\psfig{figure=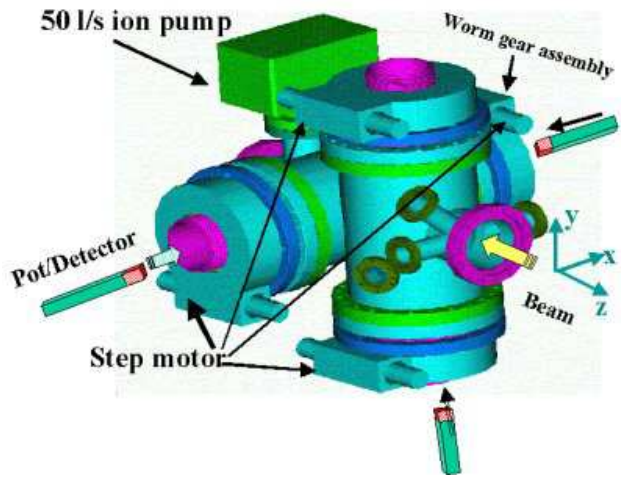,height=2.5in}}
\caption{This figure shows all main parts of the castle. The four 
``arms'' can be seen, each one allowing for a pot to be inserted. The pot 
movement is done using step motors.}
\label{fig:potes}
\end{figure}

	The design of the Roman Pots for the D\O\ FPD is shown in figure \ref{fig:potes}. They
were built by LNLS/Brazil (Laborat\'orio Nacional de Luz Synchroton) as part of 
a regional collaboration ( LAFEX/CBPF - Centro Brasileiro de Pesquisas F\'\i sicas;
UFBA - Universidade Federal da Bahia; UFRJ - Universidade Federal do Rio de Janeiro; 
UERJ - Universidade Estadual do Rio de Janeiro; IFT/UNESP - Universidade Estadual 
Paulista; and LNLS) During two years we studied many options for the castle and 
detectors. The castle was made using 316L steel following the technical
specifications to achieve a Ultra-High vacuum at the Tevatron. The combination
of four view 
quadrupole stations, as shown in figure \ref{fig:potes}, and the dipole stations, 
give the possibility to cover a large portion of the available phase space, 
allowing for a better acceptance. 
In figure \ref{fig:potes} we show the castle, indicating its main components. 
In order to have the best performance, 
the design of the pot window was submitted to a finite element analysis. 
The best results were obtained using a 150 microns foil, with elliptical cutout. 
\par 
\par	

\subsection{The Detectors}
\par
\begin{figure}[htb]
\centerline{\psfig{figure=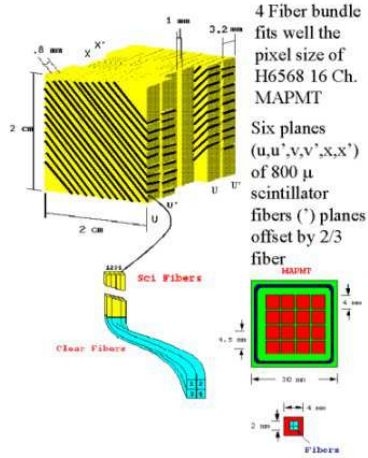,height=2.5in}}
\caption{This figure shows the six planes (u u', v v', x x') of the detector 
for the FPD spectrometer. Four scintillating fibers are 
connected together in a single channel of the multi-anode photomultiplier (MAPMT).}
\label{fig:detectors}
\end{figure}

	The FPD position detectors are based on square scintillating fibers, 800 
micron thick, as shown in figure \ref{fig:detectors}. In this figure we can see the 
frame of the scintillating fibers and 6 planes X X', U U' and V V' which compose one 
detector. The scintillating fibers are spliced (fused) to 
clear fibers which guide the signal up to the multi-anode photomultipliers (MAPMT
H6568 from Hamamatsu). There are 16 channels per X X' plane and 20 channels
for the U U' and V V' planes, giving a total of 112 channels per detector and 2016 
channels in total.
Studies about the signal, efficiency and resolution have been made. Scintillating 
fibers are the best option for our detectors among many other possible technologies.
The frame is made of polyurethane plastic. The theoretical resolution is 80 microns. 
\par 
\par
\begin{figure}[htb]
\centerline{\psfig{figure=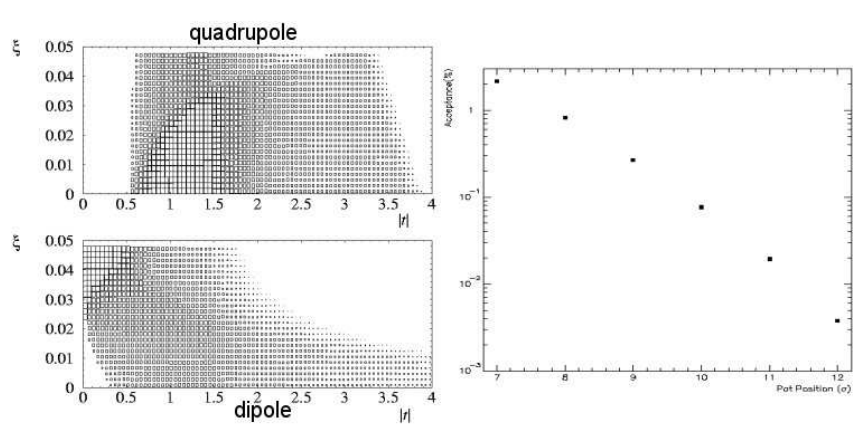,width=4.5in}}
\caption{This figure shows the $\xi$ and t acceptance for both dipole and
quadrupole stations.  The figure shows also the acceptance versus the
pot position.  We can see the acceptance variation by moving the 
pot as close as possible to the beam.}
\label{fig:acceptances}
\end{figure}

	
	The geometrical acceptance and the pot position acceptance are given in 
figure \ref{fig:acceptances}.

\section{Some Physics Topics for the Run II at Tevatron}
	The new Tevatron detectors were designed to cover a large region of 
	phase space, allowing us 
to revisit some old and interesting results, and create conditions to
observe new physics topics in diffraction. The combination of the Dipole and 
Quadrupole stations in the D\O/FPD will allow us to get a good sample of data. 
Early estimates on the size of this sample can be seen on table 
\ref{table:1}, where we also make a comparison with the available data to date.

\par

\begin{table}[ht]
\caption{This table shows a comparison between 
the present data on diffractive dijets and our 
expectation using the FPD at the Tevatron Run II.} 
\label{table:1}
\par
\begin{tabular}{||c |c |c||} \hline \hline  
Experiment    & Dijet Events  &  $E_{T}$ [GeV]  \\ \hline \hline 
     UA8         &  100       &    8            \\ \hline      
     HERA        &  Hundreds  &    5            \\ \hline          
     CDF         &  Thousands &    10           \\ \hline    
                 &  500,000   &    15           \\ \cline{2-3}  
   D\O\ /FPD     &  150,000   &    20           \\ \cline{2-3} 
                 &   15,000   &    30           \\ \hline \hline    
\end{tabular}  
\end{table}
\par

\begin{figure*}[htb]
\centerline{\psfig{figure=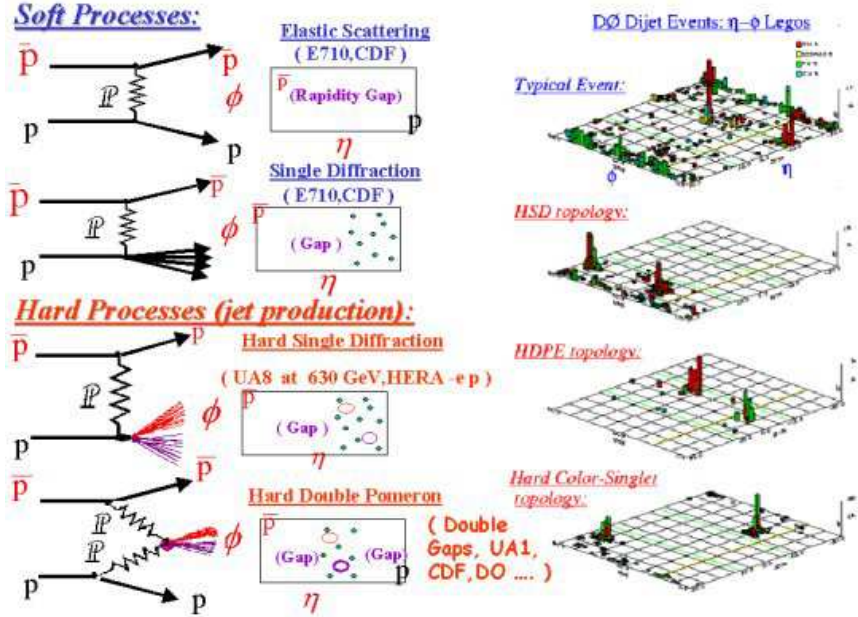,width=4.5in}}
\caption{This figure shows the possible topologies to be studied with FPD in
 the D\O\ \ detector. For each topology we have a corresponding lego plot in 
 pseudo-rapidity versus azimuthal angle. We also show the lego plots 
 corresponding to 3 topologies: the Hard Single Diffraction (HSD), the Hard Double
 Pomeron Exchange (HDPE) and the Hard Color-Singlet extracted from D\O\ \ dijet 
 events.}
\label{fig:topologies}
\end{figure*}

In figure \ref{fig:topologies} we show several topologies available for 
study using this detector, some of them, like the Hard Double Pomeron Exchange, 
have only been studied indirectly using Rapidity Gap techniques. 

\par	
\par
	We will now give a short description of some of the physics topics 
	that can be studied with the new Tevatron Detectors:
	\subsection{ Low and High $ |t| $ Elastic Scattering}
\par 
		The acceptance of the FPD spectrometers for low and high t will allow
us to extract elastic scattering events in both of these physical regions. 
The measurements of elastic cross sections can be used to extract the total 
cross section via
Optical Theorem. It is important to know the elastic slope of the differential
cross section.  The value of the slope for general differential cross sections 
characterizes a specific process which can be associated to a particular 
production (e.g. resonance production). 

\par
	\subsection{Total Cross Section}
\par
	The results from Tevatron \cite{E811} experiments are not compatible 
between them, if we extrapolate to higher energies. It would be very important 
to have another measurement at these energies. It is important to know the 
behavior of the cross sections with energy since that is related to  
the Froissart-Martin bound. After the Tevatron only the LHC will offer a new 
opportunity to make these measurements. 

\par
	\subsection{Inclusive Single Diffraction}
\par
	The inclusive single diffraction has many subtopics associated with it.
Particularly for the Tevatron detectors, the Diffractive Mass available, for
single diffraction events, $M_{x} \, = \, 450 \, GeV$, makes the extraction 
of heavy flavor physics very comfortable. Inclusive single diffraction has been a 
good laboratory for several problems in diffractive physics. In particular one
would like to study the cross section dependence on the diffractive mass
($M_{x})$ and four momentum transfered (t).
\par
	\subsection{Diffractive Jet production}
\par
	Jets have been largely studied by QCD. The discovery of diffractively 
produced jets \cite{UA8} by UA8 collaboration was very important for diffractive
physics. This was the main starting point for hard diffraction. It is expected 
that single and double diffractive jet production be exhaustively studied in the 
near future with the Tevatron and HERA detectors, making it possible to verify 
the universality of Diffractive Parton Distributions. As we see in 
table \ref{table:1} it will be possible to get dijet
events with higher transverse energies at the Tevatron Run II.
\par
	\subsection{Hard Double Pomeron Exchange}
\par
	Due to the interesting topology and exciting physics topics, double Pomeron
exchange has been largely discussed as the process for producing many different 
types of states. \cite{levin}. An advantage of the large Diffractive Mass 
available at 
the Tevatron, in this case $M_{x} \, \simeq \, 100 \, GeV$, is the possibility 
to study, by direct observation, the Pomeron $\times$ Pomeron interactions and the 
associated physics. The instrumentation available at D\O\ is appropriated to 
face the challenges of the double  Pomeron mechanism, and produce several objects 
not yet observed. Figure \ref{fig:topologies} shows the double Pomeron exchange
graph and the gaps on its corresponding lego plot. Other topics like Glueballs,
Centauros and Higgs can be exploited also in this topology.
\par 
	\subsection{Diffractive Heavy Flavor Production}
\par
	Heavy Flavor Production, has been extensively studied in high $p_t$ physics. 
Because experimental results are scarce for diffractive heavy flavor production, 
there was not enough attention to this physics. For the lack of adequate instrumentation 
we have cross sections without a clear separation of diffractive and 
non-diffractive events. In principle we can split the studies of
diffractive heavy flavor production based on the products: (i) The Charm, (ii) 
the Bottom and the (iii) Top Physics. Each one having some particularities to be 
taken into account in the diffractive production \cite{berger}.
\par
	\subsection{Diffractive W/Z boson production}
\par
	Diffractively produced W and Z bosons have been observed at the Tevatron. 
 However the present results are not satisfactory, we need more statistics for these 
kind of events, to help us understand and set constraints on the quark and gluon 
contents of the Pomeron. Both CDF \cite{cdf} and  D\O\ have made progress, and the 
current results are motivating the collaborations to proceed with this measurements. 
\par
	\subsection{Diffractive Structure Functions}
\par
	The study of Diffractive Structure Functions at the Tevatron allows 
a comparison with the existing HERA results. To understand the structure of the 
Pomeron one must know its structure function. This type of study has to be 
pursued exhaustively to get better and accurate measurements, building a clear
interpretation of the Pomeron. One should be able to answer how important are 
the gluon and the quark components of the Pomeron. The universality of the 
Pomeron components is also important for this interpretation. Practically all 
diffractive topics mentioned here depend on this knowledge.
\par	
	\subsection{Glueballs, Centauros and the Higgs}
\par
	Since the origin of QCD, Glueballs has been studied  by theoreticians 
and experimentalists.  However, we do not have a significant progress in this
field. Glueballs are perfectly valid states in QCD, so if we do not find them, 
there must be some unknown suppression rule for this bound states. That means 
we need dedicated experiments to search for glueballs 
without ambiguity with quark anti-quark bound states. The family of 
glueballs is large.  Table \ref{table:2} shows this family (oddballs are also 
shown). Oddballs should have the priority to be examined, due to the fact that  
they can not be mistaken by $q \bar q$ (meson) or qqq (baryon) bound states 
with the same quantum numbers. It is a common belief that Glueballs should be 
largely produced in the double Pomeron Exchange topology.
\par
\begin{table*}[ht]
\begin{center}
\caption{This table shows possible glueball state configurations with the 
mass and the quantum numbers for each one.} 
\label{table:2}
\par
\begin{tabular}{||c |c |c |c |c |c |c |c||} \hline \hline
\multicolumn{8}{||c||}{Glueballs and Oddballs} \\ \hline
          &     &      &   &     &  \multicolumn{3}{c||} {}           \\ 
 $J^{PC}$ &$(q \bar q)$&2g &3g&ODD& \multicolumn{3}{c||} {MASS (GeV)}\\ 
         \cline{6-8}
          &&&&&\cite{kaidalov} & \cite{morningstar} &\cite{Teper} \\ 
         \hline \hline
$0^{++}$ &YES  &YES   &YES&NO&$1.58$&$ 1.73\pm 0.13$&$1.74\pm 0.05$ \\ \hline         
$0^{+-}$ &NO&NO &YES&YES  &      &               &        \\ \hline     
$0^{-+}$ &YES  &YES   &YES&NO&      &               &        \\ \hline   
$0^{--}$ &NO&NO &YES&YES  &$2.56$&$2.59\pm 0.17$ &$2.37\pm0.27$ \\
         \hline 
$1^{++}$ &YES  &YES   &YES&NO&      &               &        \\ \hline 
$1^{+-}$ &YES  &NO &YES&NO&      &               &        \\ \hline 
$1^{-+}$ &NO&YES   &YES&YES  &      &               &        \\ \hline   
$1^{--}$ &YES  &NO &YES&NO&$3.49$&$3.85\pm0.24$  &        \\ \hline 
$2^{++}$ &YES  &YES   &YES&NO&$2.59$&$2.40\pm0.15$  &$2.47\pm 0.08$\\ 
         \hline 
$2^{+-}$ &NO&NO &YES&YES  &      &               &        \\ \hline 
$2^{-+}$ &YES  &YES   &YES&NO&$3.03$&$3.1\pm 0.18$  &$3.37\pm 0.31$ \\ 
         \hline 
$2^{--}$ &YES  &NO &YES&NO&$3.71$&$3.93\pm 0.23$ &        \\ \hline 
$3^{++}$ &YES  &YES   &YES&NO&$3.58$&$3.69\pm 0.22$ &$4.3\pm 0.34$ \\ 
         \hline 
$3^{+-}$ &YES  &NO &YES&NO&      &               &        \\ \hline 
$3^{-+}$ &NO&YES   &YES&YES  &      &               &        \\ \hline 
$3^{--}$ &YES  &NO &YES&NO&$4.03$&$4.13\pm 0.29$ &        \\ \hline 
          \hline 
\end{tabular}
\end{center}
\end{table*}
\par
\par
	Another topic is the production of Centauros, which were never 
observed in accelerator experiments. These objects were discovered in 
Cosmic Ray Physics as events with several unusual characteristics, like the 
production of a large multiplicity of charged particles, accompanied by very few 
photons. For example, as many as 100 charged particles and no more than 3 
$\pi^{0}$. \cite{halzen} There is enough center of mass energy at the Tevatron 
to produce Centauros, and since the diffractive mass is high enough, it is 
possible to produce them diffractively. The good calorimetry of the D\O\ 
detector can be very useful in observing this kind of events.

\par
	Higgs is one of the most exciting subjects for the Tevatron Run II. It  
has not been excluded the possibility that it can be produced also diffractively. We 
have two recent studies given by reference \cite{levin}, showing the 
possibility of Higgs production by the double Pomeron mechanism.

\par

\section{Conclusions}
	There is a lot of exciting new results coming from the investigation of 
diffractive phenomena, in particular the ongoing Run II at Tevatron should give 
us a better picture of diffraction with the new detectors at CDF and D\O.
Practically all possible subjects listed here can be studied with the data 
obtained from these detectors. 
\par
The main goal of the experiments in diffractive physics is hard diffraction.
Many theoretical and phenomenological progress have been made with the precise 
data from HERA, but a corresponding set of data from the Tevatron is still 
missing to allow comparisons between the two regimes of production.
This set of data will give us the opportunity to test assumptions like the 
universality of the Pomeron picture and the Diffractive Parton Distributions. 
\par
Another important topic concerns the measurement of the Total Cross Section at 
the highest energies of the Tevatron, for which there are conflicting results 
that can be settled by a new measurement from the FPD at D\O. 
This new measurement will also help to reduce uncertainties on the luminosity 
for all D\O\ physics processes. 
\par
The double Pomeron exchange and the physics that can be done with this topology 
is one of the very important subjects of this run at the Tevatron. It has not 
been excluded, as we called attention, that Centauros and Higgs \cite{levin} 
can be produced diffractively in this topology. 
\par
	Finally the diffractive physics results from CDF and D\O\ will be 
very important for future projects at the LHC, since it is expected that the 
diffractive contribution becomes more important at increasing energies.
\par
	I would like to thank the organizing committee for the superb job in 
the organization a very enjoyable meeting. In particular I would like to thank 
Profs. J. Trampetic and J. Wess for the kind invitation and financial support. 

\par

\end{document}